\documentclass[aps,showpacs,preprintnumbers,nofootinbib]{revtex4}
\usepackage{amssymb}
\usepackage{amsfonts}
\usepackage{amsmath}
\usepackage{graphicx}
\usepackage{dcolumn}
\usepackage{bm}
\usepackage[latin1]{inputenc}
\usepackage[dvips]{hyperref}
\usepackage{graphicx}

\def\be{\begin{equation}}
\def\ee{\end{equation}}
\def\bea{\begin{eqnarray}}
\def\eea{\end{eqnarray}}

\begin{document}

\title{Dirac quasinormal modes for a $4$-dimensional Lifshitz black hole}
\author{Marcela Catal\'{a}n}
\email{marceicha@gmail.com}
\affiliation{Departamento de Ciencias F\'{\i}sicas, Facultad de Ingenier\'{\i}a y Ciencias, Universidad de La Frontera, Avenida Francisco Salazar
01145, Casilla 54-D, Temuco, Chile.}
\author{Eduardo Cisternas}
\email{eduardo.cisternas@ufrontera.cl}
\affiliation{Departamento de Ciencias F\'{\i}sicas, Facultad de Ingenier\'{\i}a y Ciencias, Universidad de La Frontera, Avenida Francisco Salazar
01145, Casilla 54-D, Temuco, Chile.}
\author{P.~A.~Gonz\'{a}lez}
\email{pablo.gonzalez@udp.cl}
\affiliation{Facultad de Ingenier\'{\i}a, Universidad Diego Portales, Avenida Ej\'{e}%
rcito Libertador 441, Casilla 298-V, Santiago, Chile.}
\author{Yerko V\'{a}squez}
\email{yerko.vasquez@ufrontera.cl}
\affiliation{Departamento de F\'{\i}sica, Facultad de Ciencias, Universidad de La Serena,\\ 
Avenida Cisternas 1200, La Serena, Chile.}

\date{\today }

\begin{abstract}
We study the quasinormal modes of fermionic
perturbations for an asymptotically Lifshitz black hole in $4$-dimensions with
dynamical exponent $z=2$ and plane topology for the transverse section, and we
find analytically and numerically the quasinormal modes for massless fermionic fields by using the improved asymptotic iteration method and the Horowitz-Hubeny method. The quasinormal frequencies are purely imaginary and negative, which guarantees the stability of these black holes under massless fermionic field perturbations. Remarkably, both numerical methods yield consistent results; i.e.,
both methods converge to the exact quasinormal frequencies; however, the improved asymptotic iteration method converges in a fewer number of iterations. Also, we find analytically the quasinormal modes for massive fermionic fields for the mode with lowest angular momentum. In this case, the quasinormal frequencies are purely imaginary and negative, which guarantees the stability of these black holes under fermionic field perturbations. Moreover, we show that the lowest quasinormal frequencies have real and imaginary parts for the mode with higher angular momentum by using the improved  asymptotic iteration method.  
\end{abstract}

\maketitle


\section{Introduction}

The Lifshitz spacetimes have received great attention from the condensed
matter point of view, i.e., the searching for gravity duals of Lifshitz fixed
points due to the AdS/CFT correspondence for
condensed matter physics and quantum chromodynamics \cite{Kachru:2008yh}. From the quantum field theory point of view, there are
many invariant scale theories of interest when studying such
critical points. Such theories exhibit the anisotropic scale
invariance $t\rightarrow \lambda^zt$, $x\rightarrow \lambda x$, with $z\neq
1 $, where $z$ is the relative scale dimension of time and space, and they
are of particular interest in studies of critical exponent theory and
phase transitions. Systems with such a behavior appear, for instance, in the
description of strongly correlated electrons. The importance of possessing a
tool to study strongly-correlated condensed matter systems is beyond question, and consequently much attention has been focused on this area in recent years. Thermodynamically, it is difficult to compute conserved quantities for Lifshitz black holes; however,
progress was made on the computation of mass and related thermodynamic quantities by
using the ADT method \cite{Devecioglu:2010sf}, \cite{Devecioglu:2011yi} and
the Euclidean action approach \cite{Gonzalez:2011nz, Myung:2012cb}. Also,
phase transitions between Lifshitz black holes and other configurations with
different asymptotes have been studied in \cite{Myung:2012xc}. However, due
to their different asymptotes these phases transitions do not occur.

An important property of black holes is their quasinormal modes (QNMs) and
their quasinormal frequencies (QNFs) \cite{Regge:1957td, Zerilli:1971wd,
Zerilli:1970se, Kokkotas:1999bd, Nollert:1999ji, Konoplya:2011qq}. 
The oscillation frequency of these modes is independent of the initial
conditions and it only depends on the parameters of the black hole (mass, charge and angular
momentum) and the fundamental constants (Newton constant and cosmological
constant) that describe a black hole, just like the parameters that define
the test field. The study of the QNFs gives information about the stability
of black holes under matter fields that evolve perturbatively in their
exterior region, without backreacting on the metric. 
In general, the oscillation frequencies are complex, where the real part
represents the oscillation frequency and the imaginary part describes the
rate at which this oscillation is damped, with the stability of the black
hole being guaranteed if the imaginary part is negative. The QNFs have been
calculated by means of numerical and analytical techniques, and the
Mashhoon method, Chandrasekhar-Detweiler, WKB method, Frobenius method,
method of continued fractions, Nollert and asymptotic iteration method (AIM)
are some remarkably numerical methods. For a review see \cite{Konoplya:2011qq}
and the references therein. Generally, the Lifshitz black holes are stable
under scalar perturbations, and the QNFs show the absence of
a real part \cite{CuadrosMelgar:2011up, Gonzalez:2012de, Gonzalez:2012xc,
Myung:2012cb, Becar:2012bj,Giacomini:2012hg}. In the context of black hole
thermodynamics, the QNMs allow the quantum area spectrum of the black hole
horizon to be studied \cite{CuadrosMelgar:2011up} as well as the mass and
the entropy spectrum.

On the other hand, the QNMs determine how fast a thermal state in the
boundary theory will reach thermal equilibrium according to the AdS/CFT
correspondence \cite{Maldacena:1997re}, where the relaxation time of a
thermal state of the boundary thermal theory is proportional to the inverse
of the imaginary part of the QNFs of the dual gravity background, which was established due to the QNFs of the black hole
being related to the poles of the retarded correlation function of the
corresponding perturbations of the dual conformal field theory \cite%
{Birmingham:2001pj}. Fermions on Lifshitz Background have been studied in \cite{Alishahiha:2012nm},
by using the fermionic Green's function in $4$-dimensional Lifshitz spacetime with $z=2$, also the authors considered a non-relativistic (mixed) boundary condition for fermions and showed that the spectrum has a flat band.

In this work, we will consider a matter distribution outside the horizon of
the Lifshitz black hole in $4$-dimensions with a plane transverse section and
dynamical exponent $z=2$. The matter is parameterized by a fermionic field,
which we will perturb by assuming that there is no back reaction on the
metric. We obtain 
analytically and numerically the QNFs for massless fermionic fields
by using the improved AIM \cite{Cho:2009cj, Cho:2011sf} and the Horowitz-Hubeny method \cite{Horowitz:1999jd}, and then we study their stability under fermionic perturbations. Also, we obtain analytically the QNFs of massive fermionic fields perturbations for the mode with lowest angular momentum and numerically the lowest QNF for the mode with higher angular momentum by using the improved AIM. 

The paper is organized as follows. In Sec. II we give a brief review of the
Lifshitz black holes considered in this work. In Sec. III we calculate the
QNFs of fermionic perturbations for the $4$-dimensional Lifshitz black hole
with plane topology and $z=2$. Finally, our conclusions are in Sec. IV.

\section{Lifshitz black hole}

The Lifshitz spacetimes are described by the metrics
\begin{equation}
ds^2=- \frac{r^{2z}}{l^{2z}}dt^2+\frac{l^2}{r^2}dr^2+\frac{r^2}{l^2} d\vec{x}%
^2~,  \label{lif1}
\end{equation}
where $\vec{x}$ represents a $D-2$ dimensional spatial vector, $D$ is the
spacetime dimension and $l$ denotes the length scale in the geometry. As
mentioned, this spacetime is interesting due to it being invariant under
anisotropic scale transformation and represents the gravitational dual of
strange metals \cite{Hartnoll:2009ns}. If $z=1$, then the spacetime is the usual
anti-de Sitter metric in Poincar\'{e} coordinates. Furthermore, all scalar
curvature invariants are constant and these spacetimes have a null curvature
singularity at $r\rightarrow 0$ for $z\neq 1$, which can be seen by computing
the tidal forces between infalling particles. This singularity is reached in
finite proper time by infalling observers, so the spacetime is geodesically
incomplete, \cite{Horowitz:2011gh}. The metrics of the Lifshitz black hole asymptotically have the form (\ref{lif1}).
However, obtaining analytic
solutions does not seem to be a trivial task, and therefore constructing finite
temperature gravity duals requires the introduction of strange matter
content the theoretical motivation of which is not clear. Another way of finding
such a Lifshitz black hole solution is considering carefully-tuned
higher-curvature modifications to the Hilbert-Einstein action, as in New
Massive Gravity (NMG) in $3$-dimensions or $R^2$ corrections to General
Relativity. This has been done, for instance, in \cite{AyonBeato:2009nh,
Cai:2009ac, AyonBeato:2010tm, Dehghani:2010kd}. A $4$-dimensional topological
black hole with a hyperbolic horizon and $z=2$ was found in \cite%
{Mann:2009yx} and a set of analytical Lifshitz black holes in higher
dimensions for arbitrary $z$ in \cite{Bertoldi:2009vn}.

In this work we will consider a matter distribution outside the horizon of a
black hole that asymptotically approaches the Lifshitz spacetime with $z=2$~%
\cite{Balasubramanian:2009rx}, which is the solution for an action that
corresponds to a black hole in a system with a strongly-coupled scalar, which
is given by 
\begin{equation}
S=\frac{1}{2}\int {d^{4}x(R-2\Lambda )}-\int {d^{4}x\left( \frac{e^{-2\phi }%
}{4}F^{2}+\frac{m_{A}^{2}}{2}A^{2}+(e^{-2\phi }-1)\right) }~,
\end{equation}%
where, $R$ is the Ricci scalar, $\Lambda $ is the cosmological constant, $A$
is a gauge field, $F$ is the field strength, $m_{A}^{2}=2z$, and 
\begin{eqnarray}
\phi &=&-\frac{1}{2}log\left( 1+\frac{\rho ^{2}}{\rho _{H}^{2}}\right) ~,~A=%
\frac{f(\rho )}{\rho ^{2}}dt~,  \notag \\
ds^{2} &=&-f(\rho )\frac{dt^{2}}{\rho ^{4}}+\frac{d\rho ^{2}}{f(\rho )\rho
^{2}}+\frac{d\vec{x}^{2}}{\rho ^{2}}~,
\end{eqnarray}%
with 
\begin{equation}
f(\rho )=1-\frac{\rho ^{2}}{\rho _{H}^{2}}~.
\end{equation}
Note that the boundary of the spacetime is located at $\rho =0$. Making the
change of variable $r=1/\rho $ the metric can be put in the form 
\begin{equation}
ds^{2}=-\frac{r^{2}}{l^{2}}f\left( r\right) dt^{2}+\frac{dr^{2}}{f\left(
r\right) }+r^{2}d\vec{x}^{2}~,  \label{metric}
\end{equation}
where 
\begin{equation}
f\left( r\right) =\frac{r^{2}}{l^{2}}-\frac{1}{2}~,
\end{equation}
and by means of the change of coordinates $r=\frac{l}{\sqrt{2}}\cosh \rho $
the metric (\ref{metric}) becomes 
\begin{equation}
ds^{2}=-\frac{1}{4}\sinh ^{2}\rho \cosh ^{2}\rho dt^{2}+l^{2}d\rho ^{2}+%
\frac{l^{2}}{2}\cosh ^{2}\rho d\vec{x}^{2}~.
\end{equation}
In the next section, we will determine the QNFs by considering the Dirac
equation in this background and by establishing the boundary conditions on the fermionic field at
the horizon and at infinity.

\section{Fermionic quasinormal modes of a $4$-dimensional Lifshitz Black Hole}
A minimally coupled fermionic field to curvature in the background of a
$4$-dimensional Lifshitz Black Hole is given by the Dirac equation in curved space
\begin{equation}
\left( \gamma ^{\mu }\nabla _{\mu }+m\right) \psi =0~,
\end{equation}%
where the covariant derivative is defined as 
\begin{equation}
\nabla _{\mu }=\partial _{\mu }+\frac{1}{2}\omega _{\text{ \ \ \ }\mu
}^{ab}J_{ab}~,
\end{equation}%
and the generators of the Lorentz group $J_{ab}$ are 
\begin{equation}
J_{ab}=\frac{1}{4}\left[ \gamma _{a},\gamma _{b}\right] ~.
\end{equation}%
The gamma matrices in curved spacetime $\gamma ^{\mu }$ are defined by 
\begin{equation}
\gamma ^{\mu }=e_{\text{ \ }a}^{\mu }\gamma ^{a}~,
\end{equation}%
where $\gamma ^{a}$ are the gamma matrices in a flat spacetime. In order to
solve the Dirac equation, we use the diagonal vielbein 
\begin{equation}
e^{0}=\frac{1}{4}\sinh 2\rho dt~,\text{ \ }e^{1}=ld\rho ~,\text{ \ }e^{m}=%
\frac{l}{\sqrt{2}}\cosh \rho \tilde{e}^{m}~,
\end{equation}%
where $\tilde{e}^{m}$ denotes a vielbein for the flat base manifold $\sigma
_{\gamma }$. From the null torsion condition 
\begin{equation}
de^{a}+\omega _{\text{ \ }b}^{a}\wedge e^{b}=0~,
\end{equation}%
we obtain the spin connection 
\begin{equation}
\omega ^{01}=\frac{1}{2l}\cosh 2\rho dt~,\text{ \ }\omega ^{m1}=\frac{1}{%
\sqrt{2}}\sinh \rho \tilde{e}^{m}~,\text{ \ }\omega ^{mn}=\tilde{\omega}%
^{mn}~.
\end{equation}%
Now, using the following representation of the gamma matrices 
\begin{equation}
\gamma ^{0}=i\sigma ^{2}\otimes \mathbf{1}~,\text{ \ }\gamma ^{1}=\sigma
^{1}\otimes \mathbf{1}~,\text{ \ }\gamma ^{m}=\sigma ^{3}\otimes \tilde{%
\gamma}^{m}~,
\end{equation}%
where $\sigma ^{i}$ are the Pauli matrices, and $\tilde{\gamma}^{m}$ are the
Dirac matrices in the base manifold $\sigma _{\gamma }$, along with the
following ansatz for the fermionic field 
\begin{equation}
\psi =\frac{e^{-i\omega t}}{\sqrt{\sinh 2\rho }\cosh \rho }\left( 
\begin{array}{c}
\psi _{1} \\ 
\psi _{2}%
\end{array}%
\right) \otimes \varsigma ~,
\end{equation}%
where $\varsigma $ is a two-component fermion. The following equations are
thus obtained
\begin{eqnarray}
\partial _{\rho }\psi _{1}+\frac{4i\omega l}{\sinh 2\rho }\psi _{1}-\frac{%
\sqrt{2}i\kappa }{\cosh \rho }\psi _{2}+ml\psi _{2} &=&0~,  \label{diff} \\
\partial _{\rho }\psi _{2}-\frac{4i\omega l}{\sinh 2\rho }\psi _{2}+\frac{%
\sqrt{2}i\kappa }{\cosh \rho }\psi _{1}+ml\psi _{1} &=&0~,  \notag
\end{eqnarray}%
where $i\kappa $ is the eigenvalue of the Dirac operator in the base
submanifold $\sigma _{\gamma }$. In terms of the $r$ coordinate these
equations can be written as
\begin{eqnarray}
\sqrt{f(r)}\psi _{1}^{\prime }+\frac{i\omega l}{r\sqrt{f(r)}}\psi _{1}-\frac{%
i\kappa }{r}\psi _{2}+m\psi _{2} &=&0~,  \label{ecuacion} \\
\sqrt{f(r)}\psi _{2}^{\prime }-\frac{i\omega l}{r\sqrt{f(r)}}\psi _{2}+\frac{%
i\kappa }{r}\psi _{1}+m\psi _{1} &=&0~,  \notag
\end{eqnarray}
where the prime denotes the derivative with respect to $r$. In the following, we
analyze two cases separately, one is the case $\kappa=0$ and the other is $%
\kappa \neq 0$. First, we will find analytically the QNFs for the mode with the lowest
angular momentum, and for the modes with higher angular momentum we will obtain
the QNFs analytically and numerically by using the improved AIM and the Horowitz-Hubeny
approach.

\subsection{Case $\protect\kappa =0$}
The following substitutions 
\begin{equation}
\psi _{1}\pm \psi _{2}=\left( \cosh \rho \pm \sinh \rho \right) \left( \phi
_{1}\pm \phi _{2}\right)~,
\end{equation}
in Eqs. (\ref{diff}) and the change of variables $x=\tanh
^{2}2\rho $ enable us to obtain the following equations 
\begin{eqnarray}
4x^{1/2}\left( 1-x\right) \partial _{x}\phi _{1}+4i\omega lx^{-1/2}\phi
_{1}+\left( ml+1+4i\omega l\right) \phi _{2} &=&0~,  \notag  \label{system}
\\
4x^{1/2}\left( 1-x\right) \partial _{x}\phi _{2}-4i\omega lx^{-1/2}\phi
_{2}+\left( ml+1-4i\omega l\right) \phi _{1} &=&0~.
\end{eqnarray}
So, by decoupling $\phi_1$ from this system of equations and using%
\begin{equation}
\phi _{1}\left( x\right) =x^{\alpha }\left( 1-x\right) ^{\beta }F\left(
x\right) ~,
\end{equation}%
with 
\begin{equation}
\alpha =-i\omega l~,
\end{equation}%
\begin{equation}
\beta =-\frac{1}{4}\left( ml+1\right) ~,
\end{equation}%
we obtain the hypergeometric equation for $F\left( x\right) $ 
\begin{equation}
x\left( 1-x\right) F^{\prime \prime }\left( x\right) +\left( c-\left(
1+a+b\right) x\right) F^{\prime }\left( x\right) -abF\left( x\right) =0~,
\end{equation}%
and thus the solution is given by 
\begin{equation}
\phi _{1}=C_{1}x^{\alpha }\left( 1-x\right) ^{\beta }{_{2}F_{1}}\left(
a,b,c,x\right) +C_{2}x^{1/2-\alpha }\left( 1-x\right) ^{\beta }{_{2}F_{1}}%
\left( a-c+1,b-c+1,2-c,x\right) ~,
\end{equation}
which has three regular singular points at $x=0$, $x=1$ and $x=\infty $.
Here, $_{2}F_{1}(a,b,c;x)$ denotes the hypergeometric function and $C_{1}$, $C_{2}$
are integration constants and the other constants are defined as
\begin{equation}
a=\frac{1}{2}+\alpha +\beta ~,
\end{equation}%
\begin{equation}
b=\alpha +\beta ~,
\end{equation}%
\begin{equation}
c=\frac{1}{2}+2\alpha ~.
\end{equation}%
Now, imposing boundary conditions at the horizon, i.e., that there is only
ingoing modes, implies that $C_{2}=0$. Thus, the solution can be written as 
\begin{equation}
\phi _{1}\left( x\right) =C_{1}x^{\alpha }\left( 1-x\right) ^{\beta }{%
_{2}F_{1}}\left( a,b,c,x\right) ~.
\end{equation}
On the other hand, using Kummer's formula for hypergeometric functions, \cite{M. Abramowitz}, 
\begin{eqnarray}
{_{2}F_{1}}\left( a,b,c,x\right) &=&\frac{\Gamma \left( c\right) \Gamma
\left( c-a-b\right) }{\Gamma \left( c-a\right) \Gamma \left( c-b\right) }{%
_{2}F_{1}}\left( a,b,a+b-c,1-x\right) + \notag  \label{form}
 \\
&&\left( 1-x\right) ^{c-a-b}\frac{\Gamma \left( c\right) \Gamma \left(
a+b-c\right) }{\Gamma \left( a\right) \Gamma \left( b\right) }{_{2}F_{1}}%
\left( c-a,c-b,c-a-b+1,1-x\right)~,
\end{eqnarray}
the behavior of the field at the boundary $\left( x\rightarrow 1\right) $ is
given by 
\begin{equation}\label{A}
\phi _{1}\left( x\rightarrow 1\right) =C_{1}\left( 1-x\right) ^{\beta }\frac{%
\Gamma (c)\Gamma (c-a-b)}{\Gamma (c-a)\Gamma (c-b)}+C_{1}(1-x)^{-\beta }%
\frac{\Gamma (c)\Gamma (a+b-c)}{\Gamma (a)\Gamma (b)}~.
\end{equation}
Imposing that the fermionic field vanishes at spatial infinity, we obtain for $\beta
<0 $ the conditions $c-a=-n$ or $c-b=-n$, and for $\beta >0$ the conditions
are $a=-n$ or $b=-n$, where $n=0,1,2,...$. Therefore, the following sets of quasinormal modes are
obtained for $ml+1>0$ 
\begin{equation}
\omega =-\frac{i}{l}\left( n+\frac{ml+1}{4}\right)~,~\omega =-\frac{i}{l}%
\left( n+\frac{ml+3}{4}\right)~,
\end{equation}
and for $ml+1<0$ 
\begin{equation}
\omega =-\frac{i}{l}\left( n-\frac{ml+1}{4}\right)~,~\omega =-\frac{i}{l}%
\left( n-\frac{ml-1}{4}\right)~.
\end{equation}

Similarly, decoupling $\phi_2$ from the system of equations (\ref{system}), we obtain another set of quasinormal frequencies, for $ml+1>0$
\begin{equation}
\omega =-\frac{i}{l}\left( n+\frac{ml+3}{4}\right)~,~\omega =-\frac{i}{l}%
\left( n+\frac{ml+5}{4}\right)~,
\end{equation}
and for $ml+1<0$ 
\begin{equation}
\omega =-\frac{i}{l}\left( n-\frac{ml-1}{4}\right)~,~\omega =-\frac{i}{l}%
\left( n-\frac{ml-3}{4}\right)~.
\end{equation}

So, the imaginary part of the QNFs is negative, which
ensures the stability of the black hole under fermionic perturbations, at
least for $\kappa =0$. Remarkably, it is known that the scalar QNFs of the BTZ black hole under Dirichlet
boundary conditions permit to obtain only a set of QNFs, for positive masses
of the scalar field. However, there is another set of QNFs for a range of
imaginary masses which are allowed because the propagation of the scalar
field is stable, according to the Breitenlohner-Freedman limit, \cite%
{Breitenlohner:1982bm,Breitenlohner:1982jf}. This set of QNFs, just as the
former, can be obtained by requesting the flux to vanish at infinity, which
are known as Neumann boundary conditions. It is worth mentioning that for fermionic
perturbations there is no Breitenlohner-Freedman limit. However, it is
possible to consider Neumann boundary conditions because Dirichlet boundary
conditions would lead to the absence of QNFs for a range of masses, without
a physical reason for this absence, \cite{Birmingham:2001pj}. 
Here, we have
considered Dirichlet boundary conditions at infinity and we have found that these boundary conditions yields two set of Dirac QNFs for all range of masses (positive and negative) of the fermionic field in analogy with Neumann boundary condition which yields two set of frequencies for the BTZ black hole.

\subsection{Case $\protect\kappa \neq 0$}
In this section we will compute the QNFs for the case $\kappa \neq 0$. We will obtain analytical solutions for massless fermions, then we will employ two numerical methods as mentioned previously. Firstly, we will use the improved AIM and then we will compute some QNFs with the Horowitz-Hubeny method, and finally we will compare the results obtained with both methods.

\subsubsection{Analytical solution}
The change of variables $y=\left( \cosh ^{2}\rho -1\right) /\cosh ^{2}\rho $ in
Eqs. (\ref{diff}) makes it possible to write the system of equations 
\begin{eqnarray}
2y\left( 1-y\right) \partial _{y}\psi _{2}-2i\omega l\left( 1-y\right) \psi
_{2}+i\kappa \sqrt{2y\left( 1-y\right) }\psi _{1}+ml\sqrt{y}\psi _{1} &=&0~,
\label{sistema} \\
2y\left( 1-y\right) \partial _{y}\psi _{1}+2i\omega l\left( 1-y\right) \psi
_{1}-i\kappa \sqrt{2y\left( 1-y\right) }\psi _{2}+ml\sqrt{y}\psi _{2} &=&0~.
\notag
\end{eqnarray}
So, by decoupling this system of equations we can write the following
equation for $\psi _{1}\left( y\right)$ 
\begin{equation}
\psi _{1}^{\prime \prime }\left( y\right) +a\left( y\right) \psi
_{1}^{\prime }\left( y\right) +b\left( y\right) \psi _{1}\left( y\right) =0~,
\label{primera}
\end{equation}
where 
\begin{eqnarray}
a\left( y\right) &=&-\frac{ml\left( -1+3y\right) +i\sqrt{2}\left(
1-2y\right)\sqrt{\left( 1-y\right) }\kappa }{2\left(1-y\right) y\left( ml-%
i\kappa \sqrt{2\left( 1-y\right) }\right)}~, \\
b\left( y\right) &=&\frac{2\sqrt{2}\kappa ^{3}iy\left( 1-y\right)
^{3/2}-2\kappa l\left(1-y\right) \left( m\kappa y+\sqrt{2\left( 1-y\right) 
}\omega \right)}{4y^{2}\left(1-y\right) ^{2}\left( ml-%
i\kappa \sqrt{2\left( 1-y\right) }\right) }+  \notag \\
&& \frac{l^{3}m\left( -m^{2}y+4\left(1-y\right) ^{2}\omega ^{2}\right)
-il^{2}\left( -m^{2}\kappa y\sqrt{2\left( 1-y\right) }-2m\left(
-1+y^{2}\right) \omega\right)}{4y^{2}\left(1-y\right) ^{2}\left(
ml-i\kappa \sqrt{2\left( 1-y\right) }\right) } +  \notag \\
&& \frac{4\kappa \omega ^{2}il^{2}\sqrt{2\left( 1-y\right) }\left(1-y\right) ^{2}}{4y^{2}\left(1-y\right) ^{2}\left( ml-%
i\kappa \sqrt{2\left( 1-y\right) }\right) }~,
\end{eqnarray}
and the variable $y$ is restricted to the range $0<y<1$.
For $\psi_2$ we get similar expressions changing $\kappa$
for $-\kappa$ and $\omega$ for $-\omega$ in the above equations. Firstly, in order to obtain an analytical solution we will consider the case $m=0$. Thus, the functions $a(y)$ and $b(y)$ reduce to
\begin{eqnarray}
a\left( y\right) &=&\frac{\left( 1-2y\right) }{2\left(1-y\right) y}~, \\
b\left( y\right) &=&\frac{\omega l+2i\omega ^{2}l^{2}\left(1-y\right)
-\kappa ^{2}iy}{2iy^{2}\left(1-y\right) }~.
\end{eqnarray}
Now, using%
\begin{equation}
\psi _{1}\left( y\right) =y^{\alpha }\left( 1-y\right) ^{\beta }F\left(
y\right) ~,
\end{equation}%
with 
\begin{equation}
\alpha =-i\omega l~,
\end{equation}%
\begin{equation}
\beta =\frac{1}{2} ~,
\end{equation}%
we obtain the hypergeometric equation for $F\left( y\right) $ 
\begin{equation}
y\left( 1-y\right) F^{\prime \prime }\left( y\right) +\left( c-\left(
1+a+b\right) y\right) F^{\prime }\left( y\right) -abF\left( y\right) =0~,
\end{equation}%
and thus the solution is given by 
\begin{equation}
\psi _{1}=C_{1}y^{\alpha }\left( 1-y\right) ^{\beta }{_{2}F_{1}}\left(
a,b,c,y\right) +C_{2}y^{1/2-\alpha }\left( 1-y\right) ^{\beta }{_{2}F_{1}}%
\left( a-c+1,b-c+1,2-c,y\right) ~,
\end{equation}
which has three regular singular points at $y=0$, $y=1$ and $y=\infty $.
Here, $_{2}F_{1}(a,b,c;y)$ denotes the hypergeometric function and $C_{1}$, $C_{2}$
are integration constants and the other constants are defined as
\begin{equation}
a=\alpha +\beta-\sqrt{-\frac{\kappa^2}{2}-\omega^2 l^2} ~,
\end{equation}%
\begin{equation}
b=\alpha +\beta+\sqrt{-\frac{\kappa^2}{2}-\omega^2 l^2}~,
\end{equation}%
\begin{equation}
c=\frac{1}{2}+2\alpha ~.
\end{equation}%
Now, imposing boundary conditions at the horizon, i.e., that there is only
ingoing modes, implies that $C_{2}=0$. Thus, the solution can be written as 
\begin{equation}
\psi _{1}\left( y\right) =C_{1}y^{\alpha }\left( 1-y\right) ^{\beta }{%
_{2}F_{1}}\left( a,b,c,y\right) ~.
\end{equation}
On the other hand, using Kummer's formula for hypergeometric functions, Eq. (\ref{form}), the behavior of the field at the boundary $\left( y\rightarrow 1\right) $ is
given by 
\begin{equation}
\psi _{1}\left( y\rightarrow 1\right) =C_{1}\left( 1-y\right) ^{\beta }\frac{%
\Gamma (c)\Gamma (c-a-b)}{\Gamma (c-a)\Gamma (c-b)}+C_{1}(1-y)^{\frac{1}{2}-\beta }%
\frac{\Gamma (c)\Gamma (a+b-c)}{\Gamma (a)\Gamma (b)}~.
\end{equation}
Imposing that the fermionic field vanishes at spatial infinity, we obtain the conditions $a=-n$ or $b=-n$, where $n=0,1,2,...$. Therefore, the following set of QNFs is
obtained
\begin{equation}
\omega =-\frac{i\left( 1+4n+4n^2+2\kappa^2 \right)}{4l\left( 1+2n\right)}~.
\end{equation}
Similarly, from the equation of $\psi_2$ we obtain another set of QNFs
\begin{equation}
\omega =-\frac{i\left( 2+4n+2n^2+\kappa^2 \right)}{4l\left( 1+n\right)}~.
\end{equation}
Therefore, the imaginary part of the QNFs are negative, which
ensures the stability of the black hole under fermionic perturbations.

\subsubsection{Improved asymptotic iteration method}
In this section we will employ the improved asymptotic iteration method, which is an improved version of the method proposed in Refs. \cite{Ciftci}, \cite{Ciftci:2005xn}.  In order to apply this method, we must consider a fermionic
field by incorporating its behavior at the horizon and at infinity. Accordingly, at the horizon, $y\rightarrow 0$, the behavior of the fermionic
field is given by the solution for the fields of Eq. (\ref{primera}) at the horizon, which is 
\begin{equation}
\psi _{1}\left( y\rightarrow 0\right) \sim C_{1}y^{-i\omega
l}+C_{2}y^{1/2+i\omega l}~,
\end{equation}
\begin{equation}
\psi _{2}\left( y\rightarrow 0\right) \sim C_{1}y^{1/2-i\omega
l}+C_{2}y^{i\omega l}~.
\end{equation}
So, in order to have only ingoing waves at the horizon, we impose $C_{2}=0$, for $\psi_{1}$ and $\psi_{2}$. 
Asymptotically, from Eq. (\ref{primera}), the fermionic field behaves as 
\begin{equation}
\psi_{1} \left( y\rightarrow 1\right) \sim D_{1}\left( 1-y\right)
^{ml/2}+D_{2}\left( 1-y\right) ^{-ml/2}~.
\end{equation}
So, in order to have a regular fermionic field at infinity we impose $D_{2}=0 $ for $ml>0$. For $\psi _{2}$ the same expression is obtained. Therefore, taking into account these behaviors we define 
\begin{equation}
\psi _{1}\left( y\right) =y^{-i\omega l}\left( 1-y\right) ^{ml/2}\chi \left(
y\right)~,
\end{equation}
\begin{equation}
\psi _{2}\left( y\right) =y^{1/2-i\omega l}\left( 1-y\right) ^{ml/2}\chi
\left( y\right)~.
\end{equation}
Then, by inserting these fields in Eq. (\ref{primera}) we obtain the homogeneous
linear second-order differential equation for the function $\chi (y)$ 
\begin{equation}
\chi ^{\prime \prime }=\lambda _{0}(y)\chi ^{\prime }+s_{0}(y)\chi ~,
\label{de}
\end{equation}%
where for $\psi _{1}$
\begin{eqnarray}
\lambda _{0}(y) &=&\frac{i\kappa \left( 1-2y\right) \sqrt{2\left( 1-y\right) 
}+2l^{2}m\left( my+2i\left( 1-y\right) \omega \right) }{2y\left( 1-y\right)
\left( ml-i\kappa \sqrt{2\left( 1-y\right) }\right) }+  \notag \\
&&\frac{ml\left( -1+3y-2\kappa iy\sqrt{2\left( 1-y\right) }\right) +4\sqrt{2}%
\kappa \omega l\left( 1-y\right) ^{3/2}}{2y\left( 1-y\right) \left( ml-i\kappa \sqrt{2\left( 1-y\right) }\right) }~,
\end{eqnarray}%
\begin{eqnarray}
s_{0}(y) &=&-\frac{2\sqrt{2}i\kappa ^{3}\left( 1-y\right) ^{3/2}+ml\kappa
\left( i\sqrt{2\left( 1-y\right) }-2\kappa \left(1-y\right) \right)
-l^{3}m^{2}\left(1-y\right) \left( m-4i\omega \right) }{4y\left( 1-y\right)
^{2}\left( ml-i\kappa \sqrt{2\left( 1-y\right) }\right) }+  \notag
\\
&&\frac{ml^{2}\left(1-y\right) \left( m-im\kappa \sqrt{2\left( 1-y\right) }%
-4\kappa \omega \sqrt{2\left( 1-y\right) }\right) }{4y\left( 1-y\right)
^{2}\left( ml-i\kappa \sqrt{2\left( 1-y\right) }\right)}~,
\end{eqnarray}
and for $\psi _{2}$ we get
\begin{eqnarray}
\lambda _{0}(y) &=&\frac{i\kappa \left( -3+4y\right) \sqrt{2\left(
1-y\right) }+2l^{2}m\left( my+2i\left( 1-y\right) \omega \right) }{2y\left(
1-y\right) \left( ml+i\kappa \sqrt{2\left( 1-y\right) }\right) }+ 
\notag \\
&&\frac{ml\left( -3+5y+2\kappa iy\sqrt{2\left( 1-y\right) }\right) -4\sqrt{2}%
\kappa \omega l\left( 1-y\right) ^{3/2}}{2y\left( 1-y\right) \left( ml+%
i\kappa \sqrt{2\left( 1-y\right) }\right) }~,
\end{eqnarray}%
\begin{eqnarray}
\nonumber s_{0}(y) &=&\frac{\left( \sqrt{2}i\kappa \left( 1-y\right) ^{3/2}\left(
1+2\kappa ^{2}\right) +l^{3}m^{2}\left( 1-y\right) \left( m-4i\omega \right)
+4\sqrt{2}\left( 1-y\right) ^{3/2}\kappa l\omega \right) }{4\left(
1-y\right) ^{2}y\left( ml+i\kappa \sqrt{2\left( 1-y\right) }\right) }+ \\
\nonumber &&\frac{ml^{2}\left( 1-y\right) \left( 3m+\sqrt{2\left( 1-y\right) }im\kappa
-8i\omega +4\sqrt{2\left( 1-y\right) }\kappa \omega \right) }{4\left(
1-y\right) ^{2}y\left( ml+i\kappa \sqrt{2\left( 1-y\right) }\right) }+ \\
&&\frac{ml\left( 3\kappa i\sqrt{2\left( 1-y\right) }-2\kappa iy\sqrt{2\left(
1-y\right) }+2\left( 1+\kappa ^{2}\right) \left( 1-y\right) \right) }{%
4\left( 1-y\right) ^{2}y\left( ml+i\kappa \sqrt{2\left( 1-y\right) }\right) }~.
\end{eqnarray}
Then, in order to implement the improved AIM it is necessary to differentiate Eq. (\ref{de}) $n$ times with respect to $x$,
which yields the following equation: 
\begin{equation}
\chi ^{n+2}=\lambda _{n}(y)\chi ^{\prime }+s_{n}(y)\chi~,  \label{de1}
\end{equation}%
where 
\begin{equation}
\lambda _{n}(y)=\lambda _{n-1}^{\prime }(y)+s_{n-1}(y)+\lambda
_{0}(y)\lambda _{n-1}(y)~,  \label{Ln}
\end{equation}%
\begin{equation}
s_{n}(y)=s_{n-1}^{\prime }(y)+s_{0}(y)\lambda _{n-1}(y)~.  \label{Snn}
\end{equation}%
Then, by expanding the $\lambda _{n}$ and $s_{n}$ in a Taylor series around
the point, $\xi $, at which the improved AIM is performed 
\begin{equation}
\lambda _{n}(\xi )=\sum_{i=0}^{\infty }c_{n}^{i}(y-\xi )^{i}~,
\end{equation}%
\begin{equation}
s_{n}(\xi )=\sum_{i=0}^{\infty }d_{n}^{i}(y-\xi )^{i}~,
\end{equation}%
where the $c_{n}^{i}$ and $d_{n}^{i}$ are the $i^{th}$ Taylor coefficients
of $\lambda _{n}(\xi )$ and $s_{n}(\xi )$, respectively, and by replacing
the above expansion in Eqs. (\ref{Ln}) and (\ref{Snn}) the following
set of recursion relations for the coefficients is obtained:
\begin{equation}
c_{n}^{i}=(i+1)c_{n-1}^{i+1}+d_{n-1}^{i}+%
\sum_{k=0}^{i}c_{0}^{k}c_{n-1}^{i-k}~,
\end{equation}%
\begin{equation}
d_{n}^{i}=(i+1)d_{n-1}^{i+1}+\sum_{k=0}^{i}d_{0}^{k}c_{n-1}^{i-k}~.
\end{equation}%
In this manner, the authors of the improved AIM have avoided the
derivatives that contain the AIM in \cite{Cho:2009cj, Cho:2011sf}, and
the quantization conditions, which is equivalent to imposing a termination
to the number of iterations \cite{Barakat:2006ki}, which is given by 
\begin{equation}
d_{n}^{0}c_{n-1}^{0}-d_{n-1}^{0}c_{n}^{0}=0~.
\end{equation}
We solve this numerically to find the QNFs. In Tables \ref{QNM1} and \ref{QNM2}, we show the lowest QNFs, for a massless fermionic field with $\kappa= 1, 2$ and $3$, and $l=1$. Additionally, in Table \ref{QNM} we show the lowest QNFs for the fermionic fields with different values of the mass. In this case, the lowest QNFs have real and imaginary parts. 
The results in Table \ref{QNM1} refer to $\psi_1$  and in Table \ref{QNM2} to $\psi_2$. It is worth mentioning that a number of 25 iterations was employed for the improved AIM method. We can appreciate that the imaginary part of the QNFs are negative, which ensures the stability of the $4$-dimensional Lifshitz Black Hole under fermionic perturbations and that
for the fermionic massless field the QNFs are purely imaginary. 
\begin{table}[ht]
\caption{Improved AIM. Quasinormal frequencies for $\protect\kappa= 1, 2$ and $3$, $m=0$
and $l=1$ (set 1).}
\label{QNM1}\centering
\begin{tabular}{ccccccccc}
\hline\hline
$\kappa$ & $n$ & $\omega $ & $Exact $ & $n$ & $\omega $ & $ Exact $  \\[0.5ex] \hline
$1$ & $0$ & $-0.75000i$ & $-0.75000i$ &  $4$ & $-2.55000i$ & $-2.55000i$ \\ 
${}$ & $1$ & $-1.12500i$ & $-1.12500i$ &  $5$ & $-3.04167i$ & $-3.04167i$ \\ 
${}$ & $2$ & $-1.58333i$ & $-1.58333i$ &  $6$ & $-3.53571i$ & $-3.53571i$ \\ 
${}$ & $3$ & $-2.06250i$ & $-2.06250i$ &  $7$ & $-4.03125i$ & $-4.03125i$ \\[0.5ex] \hline
$2$ & $0$ & $-1.50000i$ & $-1.50000i$ &  $4$ & $-3.16667i$ & $-3.16667i$ \\ 
${}$ & $1$ & $-1.83333i$ & $-1.83333i$ & $5$ & $-3.64285i$ & $-3.64286i$ \\ 
${}$ & $2$ & $-2.25000i$ & $-2.25000i$ & $6$ & $-4.12500i$ & $-4.12500i$  \\ 
${}$ & $3$ & $-2.70000i$ & $-2.70000i$ & $7$ & $-4.61111i$ & $-4.61111i$ \\[0.5ex] \hline
$3$ & $0$ & $-2.12500i$ & $-2.12500i$ & $4$ & $-2.95000i$ & $-2.95000i$ \\ 
${}$ & $1$ & $-2.25000i$ & $-2.25000i$ & $5$ & $-3.37500i$ & $-3.37500i$  \\ 
${}$ & $2$ & $-2.56250i$ & $-2.56250i$ & $6$ & $-3.82143i$ & $-3.82143i$ \\ 
${}$ & $3$ & $-2.75000i$ & $-2.75000i$ & $7$ & $-4.28124i$ & $-4.28125i$ \\[1ex] \hline
\end{tabular}%
\end{table}
\begin{table}[ht]
\caption{Improved AIM. Quasinormal frequencies for $\protect\kappa= 1, 2$ and $3$, $m=0$
and $l=1$ (set 2).}
\label{QNM2}\centering
\begin{tabular}{ccccccccc}
\hline\hline
$\kappa$ & $n$ & $\omega $ & $Exact $ &  $n$ & $\omega $ & $Exact$ \\[0.5ex] \hline
$1$ & $0$ & $-0.75000i$ & $-0.75000i$ & $4$ & $-2.30556i$ & $-2.30556i$  \\ 
${}$ & $1$ & $-0.91667i$ & $-0.91667i$ & $5$ & $-2.79545i$ & $-2.79545i$  \\ 
${}$ & $2$ & $-1.35000i$ & $-1.35000i$ & $6$ & $-3.28846i$ & $-3.28846i$  \\ 
${}$ & $3$ & $-1.82143i$ & $-1.82143i$ & $7$ & $-3.78333i$ & $-3.78333i$  \\[0.5ex] \hline
$2$ & $0$ & $-1.41667i$ & $-1.41667i$ & $4$ & $-2.47222i$ & $-2.47222i$  \\ 
${}$ & $1$ & $-1.65000i$ & $-1.65000i$ & $5$ & $-2.93214i$ & $-2.93182i$  \\ 
${}$ & $2$ & $-2.03571i$ & $-2.03571i$ & $6$ & $-3.40385i$ & $-3.40385i$  \\ 
${}$ & $3$ & $-2.25000i$ & $-2.25000i$ & $7$ & $-3.88333i$ & $-3.88333i$  \\[0.5ex] \hline
$3$ & $0$ & $-2.15000i$ & $-2.15000i$ & $4$ & $-3.15909i$ & $-3.15909i$  \\ 
${}$ & $1$ & $-2.25000i$ & $-2.25000i$ & $5$ & $-3.59615i$ & $-3.59615i$  \\ 
${}$ & $2$ & $-2.39286i$ & $-2.39286i$ & $6$ & $-4.05000i$ & $-4.05000i$  \\ 
${}$ & $3$ & $-2.75000i$ & $-2.75000i$ & $7$ & $-4.51471i$ & $-4.51471i$  \\[1ex] \hline
\end{tabular}%
\end{table}
\begin{table}[ht]
\caption{Improved AIM. Lowest quasinormal frequencies for $\kappa= 1$, $m=0.5, 1.0, 1.5, 2.0$ and $2.5$,
and $l=1$.}
\label{QNM}\centering
\begin{tabular}{cc}
\hline\hline
$m$ & $\omega $ \\[0.5ex] \hline
$0.5$ & $0.08970-0.76051i$ \\ [0.5ex] \hline
$1.0$ & $0.10195-0.77728i$ \\ [0.5ex] \hline
$1.5$ & $0.09063-0.83925i$ \\ [0.5ex] \hline
$2.0$ & $0.07884-0.92493i$ \\[0.5ex] \hline
$2.5$ & $0.06907-1.02306i$ \\ [0.5ex] \hline
\end{tabular}%
\end{table}

\subsubsection{Horowitz-Hubeny method}
\begin{figure}[h]
\begin{center}
\includegraphics[width=0.7\textwidth]{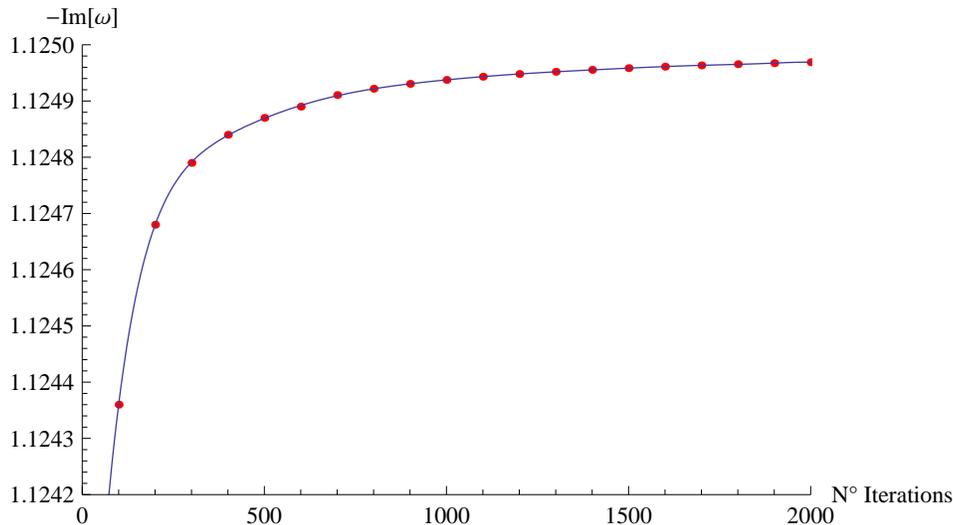}
\end{center}
\caption{The behaviour of $-Im(\omega)$ with the number of iterations for the Horowitz-Hubeny method ($n=1$, $\protect\kappa= 1$, $m=0$ and $l=1$).} \label{plots1}
\end{figure}

In this section we will employ the Horowitz-Hubeny method to evaluate some QNFs for massless fermionic field $m=0$ (for instance, see \cite{Cai:2010tr, Jing:2005ux}). In this way, we can compare both methods in order to check the results obtained in this work employing different methods. Thus, by decoupling the system of Eqs. (\ref{ecuacion}) we obtain the following equations for the fields:
\begin{equation}
\frac{ir}{2\kappa }\left( 1-\frac{2r^{2}}{l^{2}}\right) \psi _{1}^{\prime
\prime }+\frac{i}{2\kappa }\left( 1-\frac{4r^{2}}{l^{2}}\right) \psi
_{1}^{\prime }+\left( \frac{i\kappa }{r}+\frac{2r\omega }{\kappa l\left( 1-%
\frac{2r^{2}}{l^{2}}\right) }+\frac{2il^{2}\omega ^{2}}{\kappa r\left( 1-%
\frac{2r^{2}}{l^{2}}\right) }\right) \psi _{1}=0~,  \label{eqn}
\end{equation}
\begin{equation}
\frac{ir}{2\kappa }\left( 1-\frac{2r^{2}}{l^{2}}\right) \psi _{2}^{\prime
\prime }+\frac{i}{2\kappa }\left( 1-\frac{4r^{2}}{l^{2}}\right) \psi
_{2}^{\prime }+\left( \frac{i\kappa }{r}+\frac{2r\omega }{\kappa l\left( 1-%
\frac{2r^{2}}{l^{2}}\right) }-\frac{2il^{2}\omega ^{2}}{\kappa r\left( 1-%
\frac{2r^{2}}{l^{2}}\right) }\right) \psi _{2}=0~.  \label{otra}
\end{equation}
Then, by using the Tortoise coordinate
\begin{equation}
dr^{\ast }=\frac{l}{rf\left( r\right) }dr~,
\end{equation}
and also employing
\begin{equation}
\psi \left( r\right) =\left( 2r^{2}-l^{2}\right) ^{1/4}F\left( r\right)~,
\end{equation}
it is possible to write (\ref{eqn}) and (\ref{otra}) as a Schr\"{o}dinger-like equation
\begin{equation}
-\frac{d^{2}F}{dr^{\ast 2}}+V_{eff}\left( r\right) F=\omega ^{2}F~,
\end{equation}
where the effective potential $V_{eff}\left( r\right) $ is given by
\begin{equation}
V_{eff}\left( r\right) =-\frac{\kappa ^{2}}{2l^{2}}+\frac{r^{2}}{2l^{4}}+%
\frac{\kappa ^{2}r^{2}}{l^{4}}-\frac{3r^{4}}{4l^{6}}\pm \frac{ir^{2}\omega }{%
l^{3}}~.
\end{equation}
Here, the $\pm $ sign refers to two sets of QNMs associated with $\psi_1 $ and $\psi_2 $, respectively. Thus, by performing the redefinition
\begin{equation}
\psi \left( r^{\ast }\right) =e^{i\omega r^{\ast }}F\left( r^{\ast }\right)~,
\end{equation}
we get
\begin{equation}
\frac{rf\left( r\right) }{l}\frac{d^{2}\psi }{dr^{2}}+\left( \frac{d}{dr}
\left( \frac{rf\left( r\right) }{l}\right) -2i\omega \right) \frac{d\psi }{dr
}-\frac{l}{rf\left( r\right) }V_{eff}\left( r\right) \psi =0~,
\end{equation}
and with the change of coordinates $x=1/r$, this expression becomes
\begin{equation}
s\left( x\right) \frac{d^{2}\psi }{dx^{2}}+\frac{t\left( x\right) }{\left(
x-x_{+}\right) }\frac{d\psi }{dx}+\frac{u\left( x\right) }{\left(
x-x_{+}\right) ^{2}}\psi =0~,  \label{differential}
\end{equation}
where the functions $s(x)$, $t(x)$ and $u(x)$ are defined by
\begin{equation}
s\left( x\right) =-\frac{x^{2}}{2l}\left( x+x_{+}\right) ^{2}~,
\end{equation}
\begin{equation}
t\left( x\right) =x\left( x+x_{+}\right) \left( -\frac{1}{l^{3}}-\frac{x^{2}
}{2l}+2i\omega x^{2}\right)~,
\end{equation}
\begin{equation}
u\left( x\right) =x\left( -\frac{\kappa ^{2}x^{3}}{l}+\frac{x}{l^{3}}+\frac{
2\kappa ^{2}x}{l^{3}}-\frac{3}{2l^{5}x}\pm \frac{2i\omega x}{l^{2}}\right)~,
\end{equation}
and $x_{+}=\sqrt{2}/l$. The functions $s(x)$, $t(x)$ and $u(x)$ are fourth-degree polynomials. Now, we expand the polynomials around the horizon 
$x_{+}$ in the form $s\left( x\right) =\sum_{n=0}^{4}s_{n}\left(x-x_{+}\right) ^{n}$ and in a similar way for $t\left( x\right) $ and $u\left( x\right)$. Also, we expand the wave function $\psi \left( x\right) $ as
\begin{equation}
\psi \left( x\right) =\left( x-x_{+}\right) ^{\alpha }\sum_{n=0}^{\infty
}a_{n}\left( x-x_{+}\right) ^{n}~.  \label{psi}
\end{equation}
Now, in order to find the exponent $\alpha $ we have that near the event horizon the wave
function behaves as $\psi \left( x\right) =\left( x-x_{+}\right) ^{\alpha }$. So, by substituting this in Eq. (\ref{differential}) we get
\begin{equation}
\alpha \left( \alpha -1\right) s_{0}+\alpha t_{0}+u_{0}=0~,
\end{equation}
where the solutions of this algebraic equation are $\alpha =1/4$ and $\alpha
=-1/4+2i\omega l $ for the minus sign in the effective potential. The boundary
condition, i.e.  that near the event horizon there are only ingoing modes, imposes that $\alpha =1/4$.
For the plus sign the solution is $\alpha =-1/4$. Finally, by substituting $s(x)$, $t(x)$, $u(x)$ and $\psi(x)$ in (\ref{differential}) we find the following recursion relation 
\begin{equation}
a_{n}=-\frac{1}{P_{n}}\sum_{j=0}^{n-1}\left( \left( j+\alpha \right) \left(
j-1+\alpha \right) s_{n-j}+\left( j+\alpha \right) t_{n-j}+u_{n-j}\right)
a_{j}~,
\end{equation}
with
\begin{equation}
P_{n}=\left( n+\alpha \right) \left( n-1+\alpha \right) s_{0}+\left(
n+\alpha \right) t_{0}+u_{0}~.
\end{equation}
Imposing the Dirichlet boundary condition at $x\rightarrow 0$ implies that
\begin{equation}
\sum_{n=0}^{\infty }a_{n}\left( -x_{+}\right) ^{n+\alpha }=0~.
\end{equation}
Therefore,  we can obtain the QNFs solving this equation numerically. 
In Tables \ref{QNM1HH} and \ref{QNM2HH}, we show the lowest QNFs, for massless fermionic field with $\kappa= 1, 2$ and $3$, and $l=1$. 
The results in Table \ref{QNM1HH} refer to the negative sign in the effective potential ($\psi_2$) and in Table \ref{QNM2HH} to the positive sign ($\psi_1$) where we can appreciate that the imaginary part of the quasinormal
frequencies are negative, which ensures the stability of the black hole
under fermionic perturbations. It is worth mentioning that a number of $2000$ iterations was employed for the Horowitz-Hubeny method, i.e. we take up to $2000$ terms in the sum. The convergence of the quasinormal frequency with the number of iterations is shown in Figure (\ref{plots1}), for $\protect\kappa= 1$, $m=0$ and $l=1$. It is also worth mentioning that at 2000 iterations the difference between two consecutive frequencies is less than $0.000001$. Moreover, the QNFs that we have found via the Horowitz-Hubeny approach are similar to the QNFs that we found via the improved AIM, previously.
\begin{table}[ht]
\caption{Horowitz-Hubeny method. Quasinormal frequencies for $\protect\kappa= 1, 2$ and $3$, $m=0$
and $l=1$ (set 1).}
\label{QNM1HH}\centering
\begin{tabular}{ccccccccc}
\hline\hline
$\kappa$ & $n$ & $\omega $ & $Exact$ & $n$ & $\omega $ & $Exact$ \\[0.5ex] \hline
$1$ & $0$ &  $-$ & $-0.75000i$ & $4$  & $-2.54992i$ & $-2.55000i$ \\ 
${}$ & $1$ &  $-1.12497i$ & $-1.12500i$ & $5$ & $-3.04157i$ & $-3.04167i$ \\ 
${}$ & $2$ &  $-1.58328i$ & $-1.58333i$ & $6$ & $-3.53561i$ & $-3.53571i$ \\ 
${}$ & $3$ & $-2.06243i$ & $-2.06250i$ & $7$ & $-4.03114i$ & $-4.03125i$ \\[0.5ex] \hline
$2$ & $0$ &  $-1.50285i$ & $-1.50000i$ & $4$ & $-3.16662i$ & $-3.16667i$ \\ 
${}$ & $1$ &  $-1.83338i$ & $-1.83333i$ & $5$ & $-3.64279i$ & $-3.64286i$ \\ 
${}$ & $2$ & $-2.25000i$ & $-2.25000i$ & $6$ & $-4.12493i$ & $-4.12500i$ \\ 
${}$ & $3$ &  $-2.69997i$ & $-2.70000i$ & $7$ & $-4.61107i$ & $-4.61111i$ \\[0.5ex] \hline
$3$ & $0$ &  $-2.12503i$ & $-2.12500i$ & $4$ & $-2.95042i$ & $-2.95000i$ \\ 
${}$ & $1$ &  $-2.25000i$ & $-2.25000i$ & $5$ & $-3.37514i$ & $-3.37500i$ \\ 
${}$ & $2$ &  $-2.56218i$ & $-2.56250i$ & $6$ & $-3.82149i$ & $-3.82143i$ \\ 
${}$ & $3$ & $-2.75000i$ & $-2.75000i$ & $7$ & $-4.28128i$ & $-4.28125i$ \\[1ex] \hline
\end{tabular}%
\end{table}
\begin{table}[ht]
\caption{Horowitz-Hubeny method. Quasinormal frequencies for $\protect\kappa= 1, 2$ and $3$, $m=0$
and $l=1$ (set 2).}
\label{QNM2HH}\centering
\begin{tabular}{ccccccccc}
\hline\hline
$\kappa$ & $n$ & $\omega $ & $Exact$  & $n$ &  $\omega $ & $Exact$ \\[0.5ex] \hline
$1$ & $0$ &  $-0.75000i$ & $-0.75000i$ & $4$ &  $-2.30560i$ & $-2.30556i$ \\ 
${}$ & $1$ &$-0.91678i$ & $-0.91667i$ & $5$ & $-2.79549i$ & $-2.79545i$ \\ 
${}$ & $2$ & $-1.35007i$ & $-1.35000i$ & $6$ & $-3.28849i$ & $-3.28846i$ \\ 
${}$ & $3$ &$-1.82148i$ & $-1.82143i$ & $7$ &  $-3.78336$ & $-3.78333i$ \\[0.5ex] \hline
$2$ & $0$ & $-1.41669i$ & $-1.41667i$ & $4$ &  $-2.47309i$ & $-2.47222i$ \\ 
${}$ & $1$ &  $-1.64992i$ & $-1.65000i$ & $5$ &  $-2.93218i$ & $-2.93182i$ \\ 
${}$ & $2$ &  $-2.03511i$ & $-2.03571i$ & $6$ &  $-3.40409i$ & $-3.40385i$ \\ 
${}$ & $3$ &  $-2.25000i$ & $-2.25000i$ & $7$ &  $-3.88353$ & $-3.88333i$ \\[0.5ex] \hline
$3$ & $0$ & $-2.14993i$ & $-2.15000i$ & $4$ &  $-3.15893i$ & $-3.15909i$ \\ 
${}$ & $1$ & $-2.25000i$ & $-2.25000i$ & $5$ &  $-3.59577i$ & $-3.59615i$ \\ 
${}$ & $2$ &  $-2.39311i$ & $-2.39286i$ & $6$ & $-4.04918i$ & $-4.05000i$ \\ 
${}$ & $3$ & $-2.75000i$ & $-2.75000i$ & $7$ &  $-4.51185i$ & $-4.51471i$ \\[1ex] \hline
\end{tabular}%
\end{table}

\section{Conclusions}
In this work we have calculated the QNFs of massless fermionic perturbations for the
$4$-dimensional Lifshitz black hole with a plane topology and dynamical
exponent $z=2$. It is known that the boundary conditions depend on the
asymptotic behavior of spacetime. For asymptotically AdS spacetimes the
potential diverges and thus the field must be null
at infinity (Dirichlet boundary conditions) or the flux must vanish at infinity, which are known as Neumann boundary
conditions. Here, as the black hole is asymptotically Lifshitz and the potential diverges at the boundary, we have
considered that the fermionic fields will be null at infinity (Dirichlet boundary conditions) and that there are only ingoing modes at the horizon, 
and we have obtained analytical 
and numerical results using the improved AIM and the Horowit-Hubeny method, and we have found that the QNFs for the massless fermionic field are purely imaginary and negative, which ensures the stability of the black hole under massless fermionic perturbations. Remarkably, both numerical methods yield consistent results; i.e.,
both methods converge to the exact QNFs; however, the improved AIM converges in a fewer number of iterations. 

Also, we have found analytically the QNFs for massive fermionic fields for the mode with lowest angular momentum, being the QNFs purely imaginary and negative, which guarantees the stability of these black holes under fermionic fields perturbations.  Interestingly, in this case we obtain two sets of Dirac QNFs that cover all the range of mass (positive and negative) of the fermionic field in analogy with Neumann boundary condition which yields two set of modes in the BTZ black hole. On the other hand, we have shown that the lowest QNFs for massive fermionic fields for the mode with higher angular momentum, have a real and imaginary parts, by using the improved AIM. 

\section*{Acknowledgments}

P.G. would like to thank Felipe Leyton  for valuable discussions and comments on numerical methods.
This work was funded by the Comisi{\'o}n Nacional de Investigaci{\'o}n Cient{\'i}fica y Tecnol{\'o}gica through FONDECYT Grant 11121148 (YV, MC) and also partially funded by Direcci{\'o}n de investigaci{\'o}n, Universidad de La Frontera (MC). The authors also thank partial support by NLHCP (ECM-02) at CMCC UFRO. P.G. and Y.V. acknowledge the hospitality of the Universidad de La Frontera where part of this work was undertaken.
\appendix

\end{document}